\documentclass[final]{IEEEtran}
\pdfoutput=1
\usepackage{amsthm,amssymb,graphicx,multirow,amsmath,color,amsfonts}
\usepackage[update,prepend]{epstopdf}
\usepackage[noadjust]{cite}
\usepackage[latin1]{inputenc}
\usepackage{tikz}
\usepackage{bbm} 
\usepackage{pdfpages}
\usepackage{multirow}
\usepackage{comment}
\usepackage[pdftex]{hyperref}

\def\nb0{{\mathbf{0}}}
\def\nb1{{\mathbf{1}}}

\allowdisplaybreaks 

\begin{document}
\graphicspath{{./Figures/}}
\title{
On the Role of Age of Information in the Internet of Things 
}
\author{
Mohamed A. Abd-Elmagid, Nikolaos Pappas, and Harpreet S. Dhillon
\thanks{Mohamed A. Abd-Elmagid and Harpreet S. Dhillon are with Wireless@VT, Department of ECE, Virginia Tech, Blacksburg, VA (Email: \{maelaziz,\ hdhillon\}@vt.edu). Nikolaos Pappas is with the Department of Science and Technology, Link\"{o}ping University, SE-60174 Norrk\"{o}ping, Sweden (Email: nikolaos.pappas@liu.se). The support of the U.S. NSF (Grant CPS-1739642) is gratefully acknowledged.}
}


\maketitle

\begin{abstract}
In this article, we provide an accessible introduction to the emerging idea of {\em Age of Information (AoI)} that quantifies {\em freshness} of information and explore its possible role in the efficient design of freshness-aware Internet of Things (IoT). We start by summarizing the concept of AoI and its variants with emphasis on the differences between AoI and other well-known performance metrics in the literature, such as throughput and delay. Building on this, we explore freshness-aware IoT design for a network in which IoT devices sense potentially different physical processes and are supposed to frequently update the status of these processes at a destination node (such as a cellular base station). Inspired by the recent interest, we also assume that these IoT devices are powered by wireless energy transfer by the destination node. For this setting, we investigate the optimal {\em sampling policy} that jointly optimizes wireless energy transfer and scheduling of update packet transmissions from IoT devices with the goal of minimizing long-term weighted sum-AoI. Using this, we characterize the achievable {\em AoI region}. We also compare this AoI-optimal policy with the one that maximizes average throughput (throughput-optimal policy), and demonstrate the impact of {\em system state} on their structures. Several promising directions for future research are also presented.
\end{abstract}
\section{Introduction} \label{sec:intro}
The Internet of Things (IoT) is an emerging {\em digital fabric} that will tightly integrate our physical world into computer networks by connecting billions of {\em things}, such as small sensors, wearables, vehicles, and actuators, to the Internet. This global revolution is already transforming our cities and villages into smarter and more connected communities. 
An IoT network consists of three main components: i) IoT devices, ii) communication network, and iii) destination nodes. The IoT devices are usually deployed to observe some physical characteristic of the environment for a certain geographical area, e.g., temperature, pollution levels, or humidity. The sensed data measurements are transmitted through the communication network to the destination nodes where they are processed to extract meaningful information, e.g., controllable output decisions or remote source reconstruction that can assist in the prediction of its information status evolution. Clearly, the accuracy of such output decisions, which determines the performance of IoT-enabled applications, is directly related to the {\em freshness} of the aggregated data measurements of the IoT devices at the destination nodes \cite{shreedhar2018acp}. Particularly, the duration of time over which the information status at a destination node is still considered fresh is dependent on the application for which IoT devices are being used. For instance, this duration can be relatively large if the IoT devices are deployed to sense temperature or humidity, whereas it may be very small for human safety applications.

Before designing an IoT network that preserves freshness of information at the destination nodes, we need to rigorously quantify {\em information freshness}. In this article, we use the concept of Age of Information (AoI) for this purpose \cite{kaul2012real,costa2016age,Modiano2015,sun2017update,ABedewy2016,kosta2017age_mono}. 
 AoI was first proposed in \cite{kaul2012real} as a new metric that captures how frequently the information status at a destination node (also referred to as a {\em monitor}) needs to be updated through status update transmissions from a source node. In the context of IoT networks, the source node may refer to a single IoT device or an aggregator located near a group of IoT devices, which transmits measurements of sensed information to the destination node \cite{abd2018average}. The energy-constrained nature of the IoT devices along with network congestion increase the likelihood of packet loss or out of order reception, which in turn reduces the {\em value} of update packets received at the destination node and results in wastage of resources due to obsolete transmissions. It becomes even worse for far-off IoT devices whose direct links to the destination nodes may be very poor \cite{7842431}. This necessitates the need for characterizing AoI for these networks.

Due to its ubiquity and cost efficient implementation, radio-frequency (RF) energy harvesting has quickly emerged as an appealing solution for powering IoT devices, the majority of which are low power devices, such as sensors \cite{abd2018coverage}. After introducing the idea of AoI, our objective is to investigate the role of AoI in designing freshness-aware RF-powered IoT networks. 
%
Towards this objective, we first propose a generic system setup for an IoT network, in which RF-powered IoT devices are sensing different physical processes and need to transmit their sensed data to a destination node. We then investigate the optimal sampling policy for IoT devices that minimizes the long-term weighted sum-AoI. Particularly, we jointly optimize wireless energy transfer by the destination node and scheduling of update packet transmissions from IoT devices.
 Our results demonstrate that the AoI-optimal and throughput-optimal policies have completely different structures. We also characterize the achievable AoI region and demonstrate a fundamental trade-off between achieving fairness among different processes and achieving the minimum sum-AoI. To the best of our knowledge, this article makes the first attempt to efficiently design freshness-aware IoT networks while incorporating RF-energy harvesting.
 \section{Age-of-Information and its Variations}\label{sec:concept}
\begin{figure}[t!]
\centering
\includegraphics[width=0.8\columnwidth]{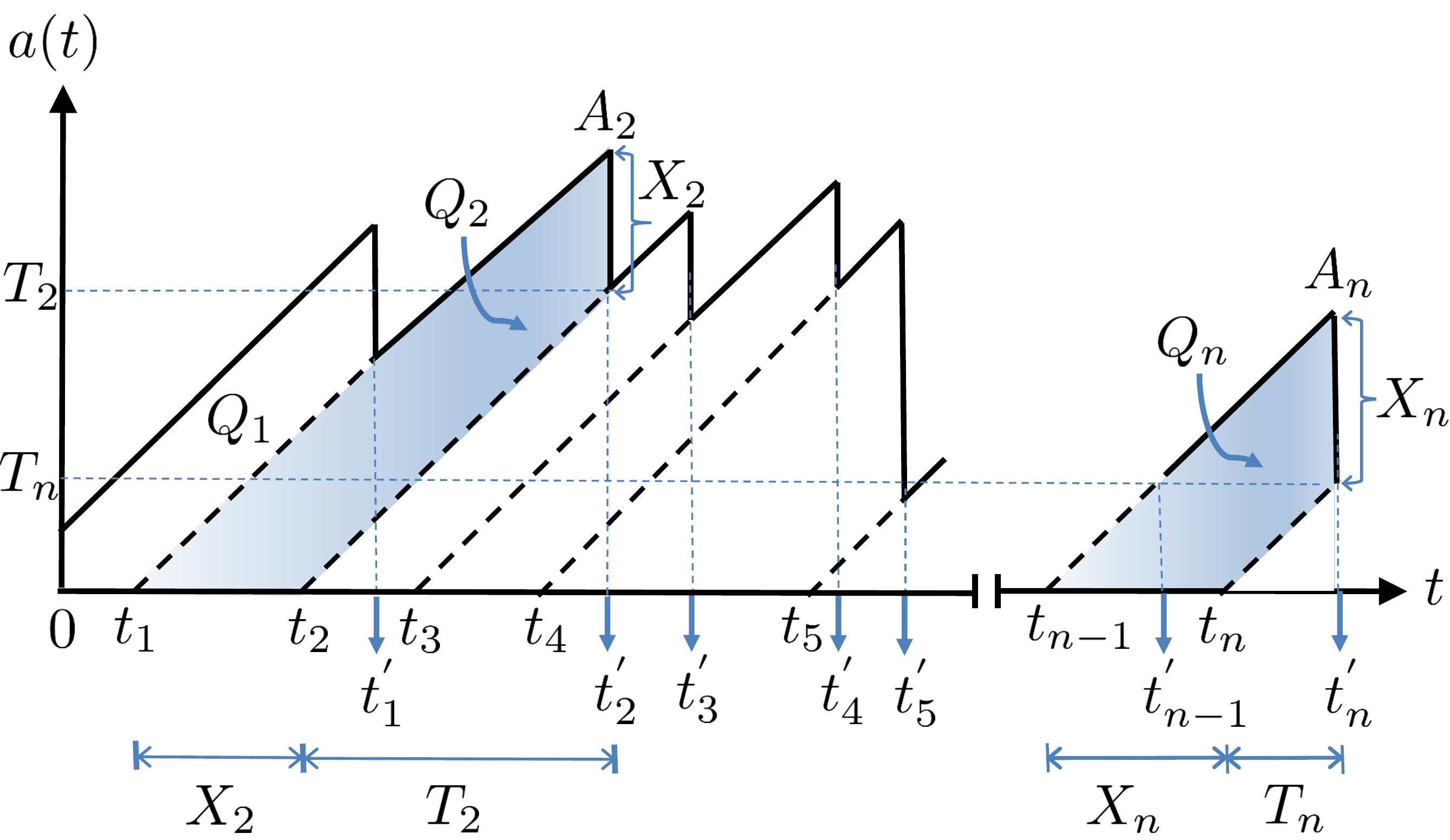}
\caption{AoI evolution vs. time for $n$ update packets.}
\label{fig:illu}
\end{figure}

 Real-time status updates are indispensable for many key applications, such as predicting and controlling forest fires, safety of an intelligent transportation system, and efficient energy consumption in future smart homes. A common setup for status update systems is the existence of a source node that generates update packets, and then transmits them through a communication system to a destination node. First introduced in \cite{kaul2012real}, AoI is a new metric that quantifies freshness of information at a monitor about some remote stochastic process observed by the source node. More formally, AoI is defined as the time elapsed since the last successfully received update packet at the monitor was generated at the source. In order to introduce the idea of AoI concretely, we use Fig. \ref{fig:illu}, which depicts a realization of AoI, denoted by $a(t)$, at the monitor as a function of time when the source transmits update packets using a First-Come-First-Served (FCFS) discipline and only one packet transmission may occur at any given time. Here, $t_n$ and $t'_n$ denote the generation and reception time instants of packet $n$ at the source and monitor, respectively. Therefore, we observe that: i) $X_n$ is the interarrival time between packets $n - 1$ and $n$, i.e., the time elapsed between the generation of packets $n - 1$ and $n$, ii) $T_n$ is the system time of packet $n$, i.e., the time elapsed from the generation of packet $n$ at the source until it is being received at the monitor, and iii) AoI is reset to $T_n$ at $t'_n$ since packet $n$ becomes the latest received update packet at $t'_n$, and hence the AoI value at that time instant is the time passed since the generation of packet $n$, which is $T_n$. 
 
Since it is not straightforward to characterize the distribution of AoI, the focus is mostly on characterizing average AoI and its variants, which have enabled many useful analytical studies. One can use a simple geometric construction for this calculation by assuming that the AoI process is ergodic for which the time average of any of its sample paths is the same as its ensemble average. Therefore, the expression for average AoI can be derived by computing the time average of any sample path, e.g., $a(t)$ depicted in Fig. \ref{fig:illu}, which comes out to be a function of the interarrival time and the system time of different update packets \cite{kaul2012real}.
 
 %
 For analytical tractability, it is sometimes useful to work with a lower bound on the average AoI obtained by ignoring the {\em waiting time} between the generation of a packet and its transmission. This is done by assuming that the source node has a {\em generate-at-will} ability and adopts a {\em just-in-time} (also referred to as {\em zero-time}) update policy \cite{kaul2012real}. Under the generate-at-will policy, the source node is capable of both observing the state of the communication channel (idle or busy) and generating update packets at any time of its own choice. Similarly, the just-in-time update policy means that a new update packet is instantaneously generated by the source node and starts its service time right after the current update packet in service is delivered to the destination node.
%
 
In the above formulation, AoI increases linearly between any two consecutive update packet receptions. In other words, AoI assumes the cost of information staleness to be directly proportional to the time elapsed since the last update packet received at the monitor, where the cost is defined in time units.
 Expanding the notion of AoI, a more generic cost of information staleness metric, namely, age penalty function (also referred to as {\em Cost of Update Delay (CoUD)} in \cite{kosta2017age}), has been recently proposed \cite{sun2017update}. The CoUD can take any form of a payment function, which is non-negative and monotonically increasing, to quantify the cost of information absence at the monitor. The type of cost function needs to be properly chosen based on the statistics of the observed stochastic process at the source. For a generate-at-will model, the optimal update policy was investigated in \cite{sun2017update}, while considering a general form for the age penalty function. Surprisingly, it was shown that the just-in-time policy 
 is only optimal when the service times of update packets are constant whereas it is not optimal for some commonly used service time distributions in queueing theory, such as exponential, geometric, Erlang, and hyperexponential.

One way to keep the information status fresh at the monitor is to minimize the average AoI. However, this may not be mathematically tractable in many situations. This has inspired some simpler measures for the information age. Next, we present two such tractable information staleness metrics:
\begin{itemize} 
\item {\em Peak Age of Information (PAoI)}. The PAoI characterizes the maximum value of AoI immediately before an update packet is received at the monitor \cite{costa2016age}. For instance, as observed in Fig. \ref{fig:illu}, the value of PAoI associated with update $n$ is $A_n$. Clearly, PAoI provides information about the worst case values of AoI, and its probability distribution can be derived relatively easily (compared to AoI) owing to its simpler structure. Therefore, the PAoI is particularly applicable when the prime goal is to maintain the worst value of AoI below a system design threshold with a certain probability.
\item {\em Value of Information of Update (VoIU)}. The VoIU quantifies the importance of the update packet received at the monitor \cite{kosta2017age}. More specifically, when the monitor receives a new update packet, its uncertainty about the current value of the observed stochastic process at the source is reduced. The importance of this newly received update is defined by how much it improves the monitor's prediction accuracy about the current status of the observed stochastic process. 
For a concrete example, consider Fig. \ref{fig:illu}, where the information age reduction due to the reception of update packet $n$ is $X_n$, and the VoIU quantifies how large this reduction is with respect to the PAoI associated with packet $n$, i.e., the VoIU associated with packet $n$ is $\dfrac{X_n}{A_n}$.
\end{itemize}

{\em Comparison with traditional metrics.} Two of the most popular design goals in communication systems are maximizing the system throughput and minimizing the end-to-end delay. The takeaway message of \cite{kaul2012real} is that, for a fixed service rate of update packets, the optimal rate at which the source should generate its update packets in order to minimize the average AoI is different from the optimal rates that either maximize throughput or minimize delay. Intuitively, maximizing the system throughput is equivalent to transmitting update packets as fast as possible. As a result, the update packets become backlogged in the communication system, and the VoIU associated with each update packet received at the monitor will be significantly reduced. On the other hand, minimizing the delay or equivalently minimizing the system time of each update packet is achieved by reducing the rate of update generations at the source. In this case, the monitor will unnecessarily have outdated status information due to the lack of update packet receptions. Unlike these well-known performance metrics, the concept of AoI allows to include contextual aspects of system design. More specifically, the transmitted packets do not have the same importance or equivalently the VoIU associated with different update packet receptions are not the same. We revisit this point in the next section in the context of freshness-aware IoT. In particular, we demonstrate the differences between the structures of the AoI-optimal and throughput-optimal polices as a function of system state variables.
\section{Age-of-Information for IoT: Network Design and Operation}\label{sec:IoT}
\begin{figure}[t!]
\centering
\includegraphics[width=1\columnwidth]{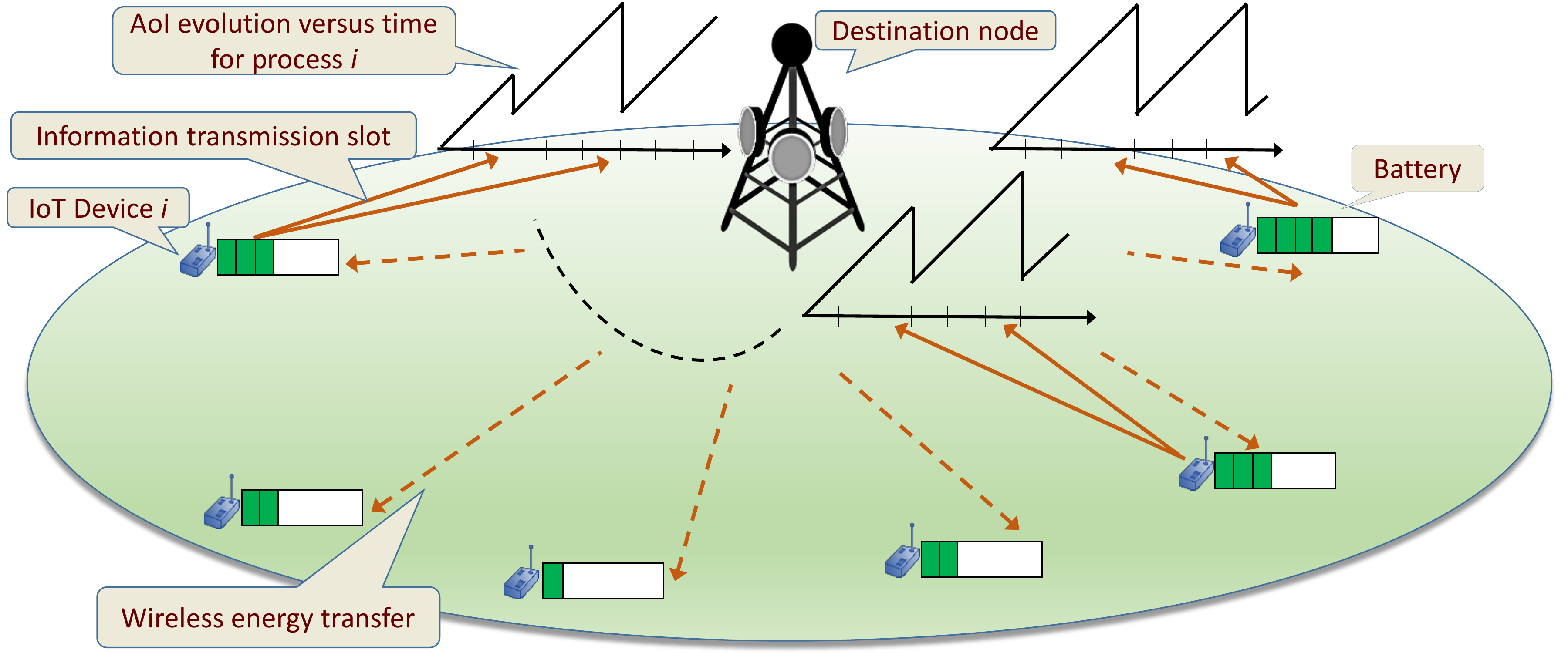}
\caption{An illustration of the system setup.}
\label{sys_model}
\end{figure}
\subsection{Network Model}
We consider an IoT network composed of a destination node (for instance, a cellular base station) and $K$ IoT devices, as shown in Fig. \ref{sys_model}. As already discussed in Section \ref{sec:intro}, each ``IoT device'' may refer to a single device (as considered in the sequel) or an aggregator located near a group of IoT devices. Each IoT device is deployed to observe some physical process (e.g., temperature, humidity, etc.) and to transmit update packets to the destination so that the information status of its observed process at the destination remains {\em fresh}. Note that this is a generalization of the single source-destination pair model considered in most prior works on AoI in the literature \cite{kaul2012real,costa2016age,sun2017update,
kosta2017age,kosta2017age_mono}.

To enable a {\it self-perpetuating} operation of the IoT network, each IoT device is equipped with an RF energy harvesting circuitry as its only source of energy. Particularly, each device harvests energy from the RF signals transmitted by the destination in the downlink, and stores it in a battery with finite capacity. The stored energy is then used for transmitting update packets to the destination. Note that the destination node is assumed to have a stable energy supply. It is assumed that all IoT devices operate over the same frequency channel and each device is equipped with a single antenna. Therefore, at a given time instant, each IoT device can either harvest wireless energy in the downlink or transmit data in the uplink. Time is assumed to be slotted with the duration of each slot being $T$ seconds. It is assumed that both downlink and uplink channels between the destination node and IoT devices are affected by quasi-static flat fading.
\subsection{State and Action Spaces}
 At the beginning of a time slot, the state of each IoT device is characterized by its battery level, its uplink and downlink channel power gains and the AoI value for its observed process at the destination node. The system state is then defined as the combination of all different states of IoT devices. In addition, the AoI value for each process at the destination node is assumed to be upper bounded by a finite value which can be chosen to be arbitrarily large \cite{zhou2018joint}. This value signifies that the information is too stale to be of any use at the destination node. 
Based on the system state, one of two potential actions is decided:
\begin{itemize}
\item {\em Information transmission}. When a time slot is dedicated for information transmission, one of the IoT devices transmits an update packet about its observed process to the destination node. We consider a generate-at-will policy, where whenever an IoT device is allocated a time slot for data transmission, it generates an update packet at the beginning of that time slot. Clearly, the choice of a given IoT device for information transmission is constrained by the availability of energy required for an update packet transmission at its battery, and the amount of this required energy mainly depends on the quality of its uplink channel and the packet size.
\item {\em Wireless energy transfer}. When a time slot is allocated for wireless energy transfer, the destination node broadcasts wireless energy in the downlink. The amount of energy harvested by each IoT device depends on the quality of its downlink channel and the efficiency of its energy harvesting circuitry. We assume that the transmit power by the destination node is sufficiently large such that the energy harvested at each IoT device due to uplink data transmissions by other devices is negligible compared to the energy it harvests from the downlink transmissions. 
\end{itemize}
\subsection{Problem Statement and System Design Insights}
Given an {\em importance weight} for each process at the destination node, we investigate the optimal strategy, which establishes the decisions taken at different states of the system, achieving the minimum weighted sum of average AoI values for different processes at the destination node.
 The problem can be formulated as a Markov Decision Process (MDP) with finite state and action spaces via discretizing the battery levels and channel power gains, and hence it could be solved using the Value Iteration Algorithm (VIA) or the Policy Iteration Algorithm (PIA).
\begin{table}[t!]\caption{Table of simulation setup}
\centering
\begin{center}
\scalebox{0.8}{
    \begin{tabular}{ {c} | {c} }
    \hline\hline
    \textbf{Parameter} & \textbf{Value} \\ \hline
        $T$ & $1$ second \\ \hline
        $\beta$ & $2$ \\ \hline
        Bandwidth & $1$ MHz \\ \hline
        Transmit power of the destination node & $37$ dBm \\ \hline
        Noise power & $- 95$ dBm \\ \hline
        Efficiency of the energy harvesting circuitry & $0.5$ \\ \hline
        Antenna gain at the destination node & $G = 7$ dB \\ \hline
        Upper bound to the AoI value & $A_{1,{\rm max}} = 10$ \\ \hline\hline
    \end{tabular}}
\end{center}
\label{tab:TableOfsimulation}
\end{table}
\begin{figure}[t!]
\centering
\includegraphics[width=0.8\columnwidth]{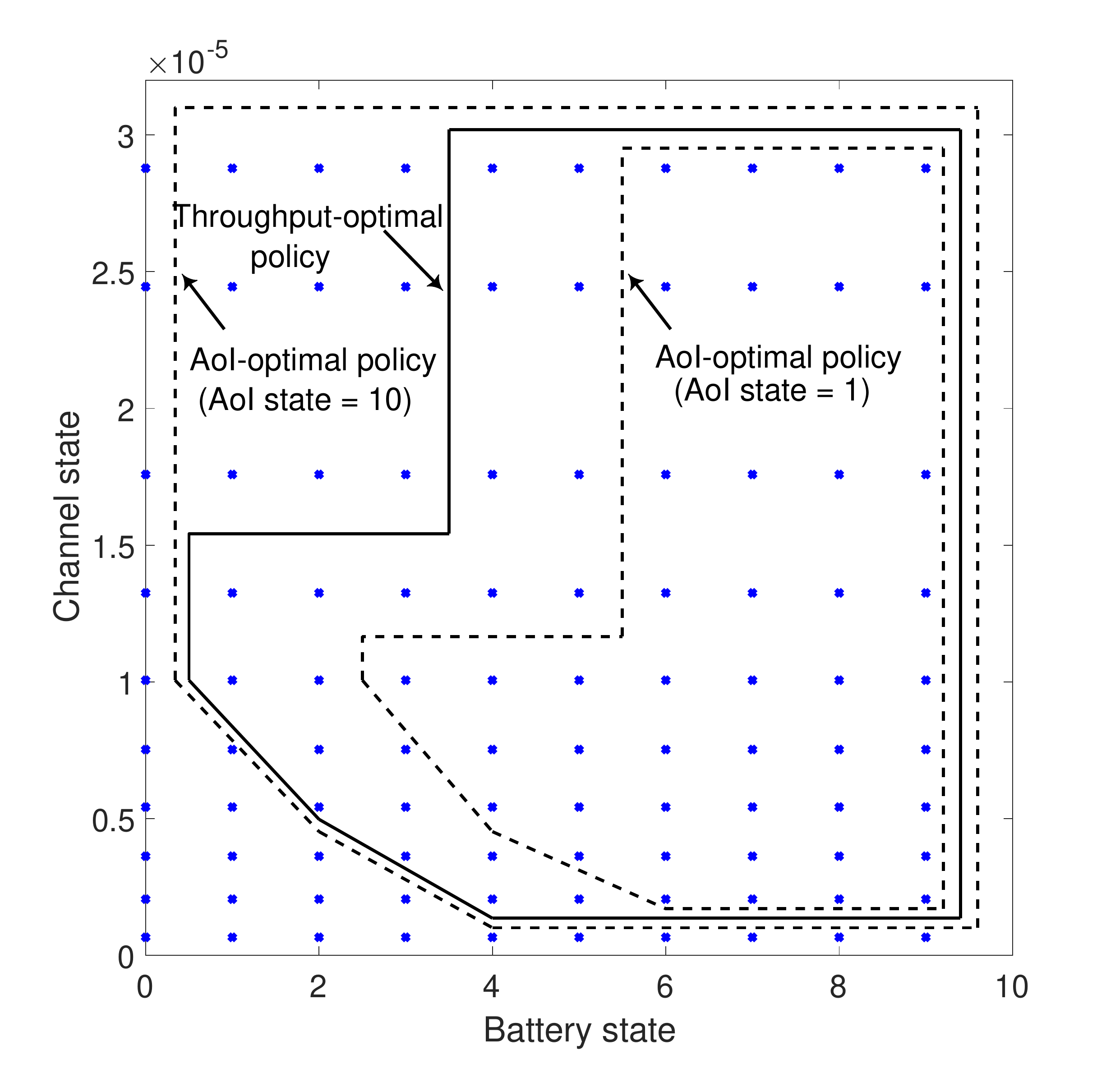}
\caption{Comparison between AoI and throughput for the case of a single IoT device. Battery capacity is $0.3$ mjoules, update packet size is $12$ Mbits and $d_1 = 40$ meters.}
\label{throughput_vs_AoI}
\end{figure}

For the classical single source-destination pair setting studied in the literature, i.e., considering the case of a single IoT device in our system setup, we first compare the structures of the AoI-optimal and throughput-optimal policies in Fig. \ref{throughput_vs_AoI}. The downlink and uplink channel power gains between IoT device $i$ and the destination node are modeled as $2 \times 10^{-2}\psi_i^2 d_i^{-\beta}$, where $\psi_i^2\sim \exp(1)$ denotes the small-scale fading gain and $d_i^{-\beta}$ represents standard power law path-loss with exponent $\beta$. 
 The default values of the system parameters are given in Table \ref{tab:TableOfsimulation}. The channel power gains and battery are discretized into $10$ levels, where each blue point in Fig. \ref{throughput_vs_AoI} represents a potential state of the system, and the battery state denotes the number of energy quanta inside the battery. Since the slot length is unity, the AoI state represents the number of time slots passed since the generation of the latest received update packet at the destination. Note that the polygons inside Fig. \ref{throughput_vs_AoI} represent the optimal solutions of the MDPs modeling the average AoI minimization problem and the average throughput maximization problem, respectively. For the combinations of battery and channel states that lie inside a given polygon, the optimal decision is to transmit an update packet. Similarly, for the combinations located outside the polygon, the optimal decision is to allocate the time slot for wireless energy transfer. For this simulation setup, the set of numbers of energy packets required for transmitting update packets with the increase of the channel state is $\{12, 4, 3, 2, 2, 1, 1, 1, 1, 1\}$. Here the number $12$ means that the IoT device can not transmit an update packet in the worst channel state even if its battery is full of energy since the capacity of battery is $9$.

In order to highlight the effect of the AoI state on the AoI-optimal policy, we compare the throughput-optimal policy, represented by the solid polygon, with the AoI-optimal policy in two different regimes: (a) when the AoI value is $1$ (represented by the inner dotted polygon), which indicates that the previous time slot was dedicated for information transmission, and (b) when the AoI value is $10$ (represented by the outer dotted polygon), i.e., AoI reaches its maximum possible value, which indicates that the information status has expired at the destination node. The key message from the AoI-optimal policy is that it is wise not to transmit an update packet when the AoI state is low as long as the battery state value is small (for instance when the battery state lies between $1$ and $5$ and the AoI state value is $1$). Instead, allocating the time slot for wireless energy transfer will help (by increasing the available energy in the battery) to transmit update packets in future slots when the value of AoI grows. On the other hand, if the AoI state is high, it is always optimal to transmit an update packet whenever the IoT device has enough energy required for information transmission. The structures of the AoI-optimal and throughput-optimal policies are different, as can be observed from the solid polygon. Although, for example, we have the ability of transmitting an update packet when the battery state value lies in the range $1 \rightarrow 3$ and the channel state is quite good, we prefer instead to allocate the time slot for wireless energy transfer and utilize the high amount of energy harvested in that good channel state for update packet transmissions in future slots. 
\begin{figure}[t!]
\centering
\includegraphics[width=0.7\columnwidth]{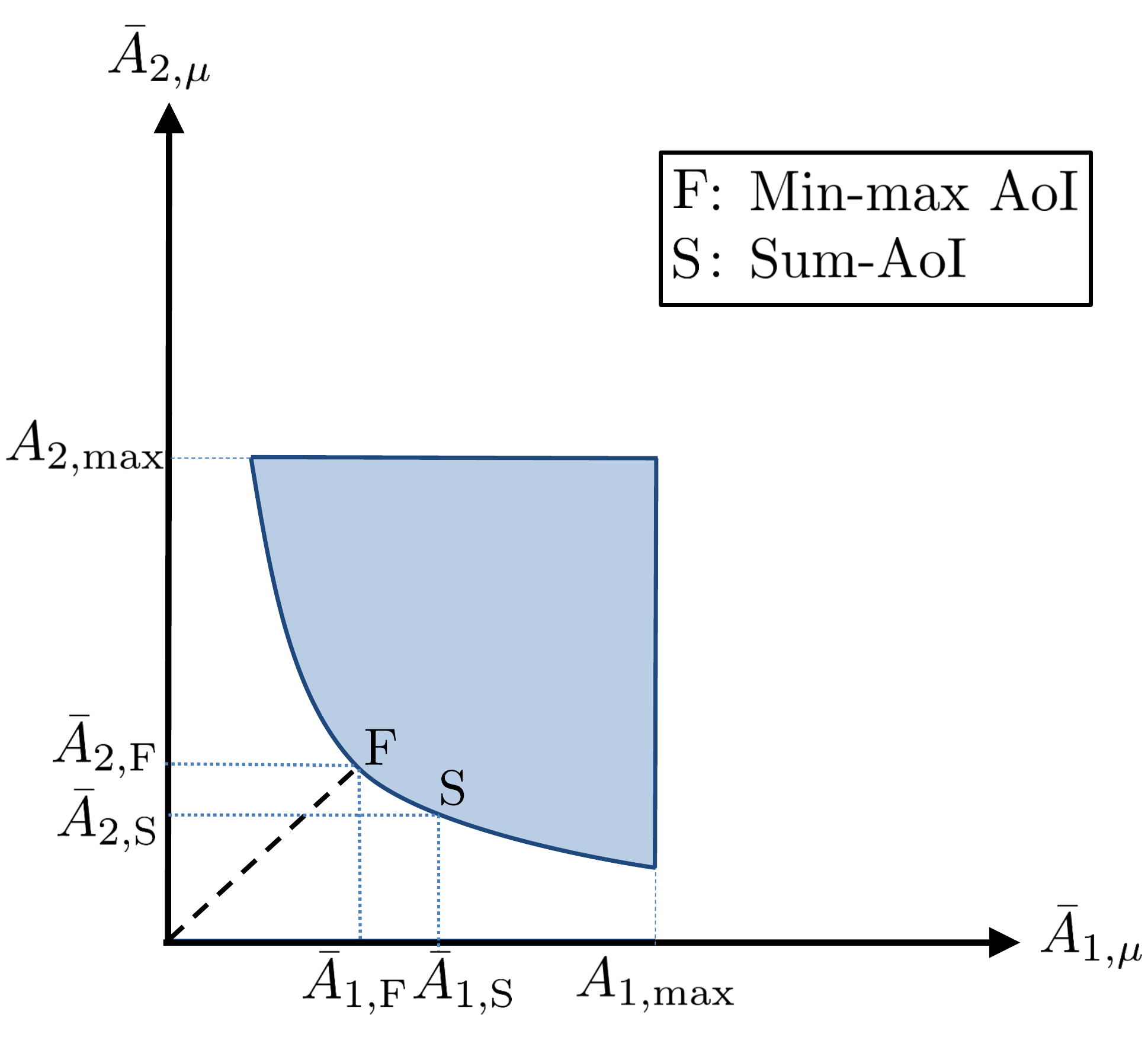}
\caption{Illustration of achievable AoI region for $K = 2$.}
\label{AoI_region}
\end{figure}

Next, we aim at characterizing the {\em achievable AoI region}. The achievable AoI region can be obtained by evaluating the optimal average AoI values of all processes for different combinations of their importance weights at the destination node. More concretely, in Fig. \ref{AoI_region}, the achievable AoI region is represented by the shaded region for the case of two IoT devices. Particularly, for different combinations of importance weights, different operating points inside the boundary of the AoI region could be achieved. 
Two operating points of particular interest are as follows. 
\begin{itemize}
\item {\em Sum-AoI}. This operating point is represented by S in Fig. \ref{AoI_region} and can be obtained by choosing the same importance weights for different processes. Note that the devices that are located closer to the destination node experience better channels in both downlink (resulting in a higher harvested energy) and uplink (thus requiring less energy for uplink transmission) than the far-off devices. Hence, the associated optimal policy with sum-AoI allocates, on average, more time slots for update packet transmissions of closer devices to the destination node, thereby making it unfair for far-off devices.
%
\item {\em Min-max AoI}. This operating point is represented by F in Fig. \ref{AoI_region} and can be obtained by choosing the values of importance weights for different processes such that the maximum average AoI among them is minimized. Equivalently, this problem could also be viewed as minimizing the achievable common AoI value for all processes. Hence, the associated optimal policy guarantees fairness between the achievable average AoI values for all processes.
\end{itemize} 
\begin{figure}[t!]
\centering
\includegraphics[width=0.8\columnwidth]{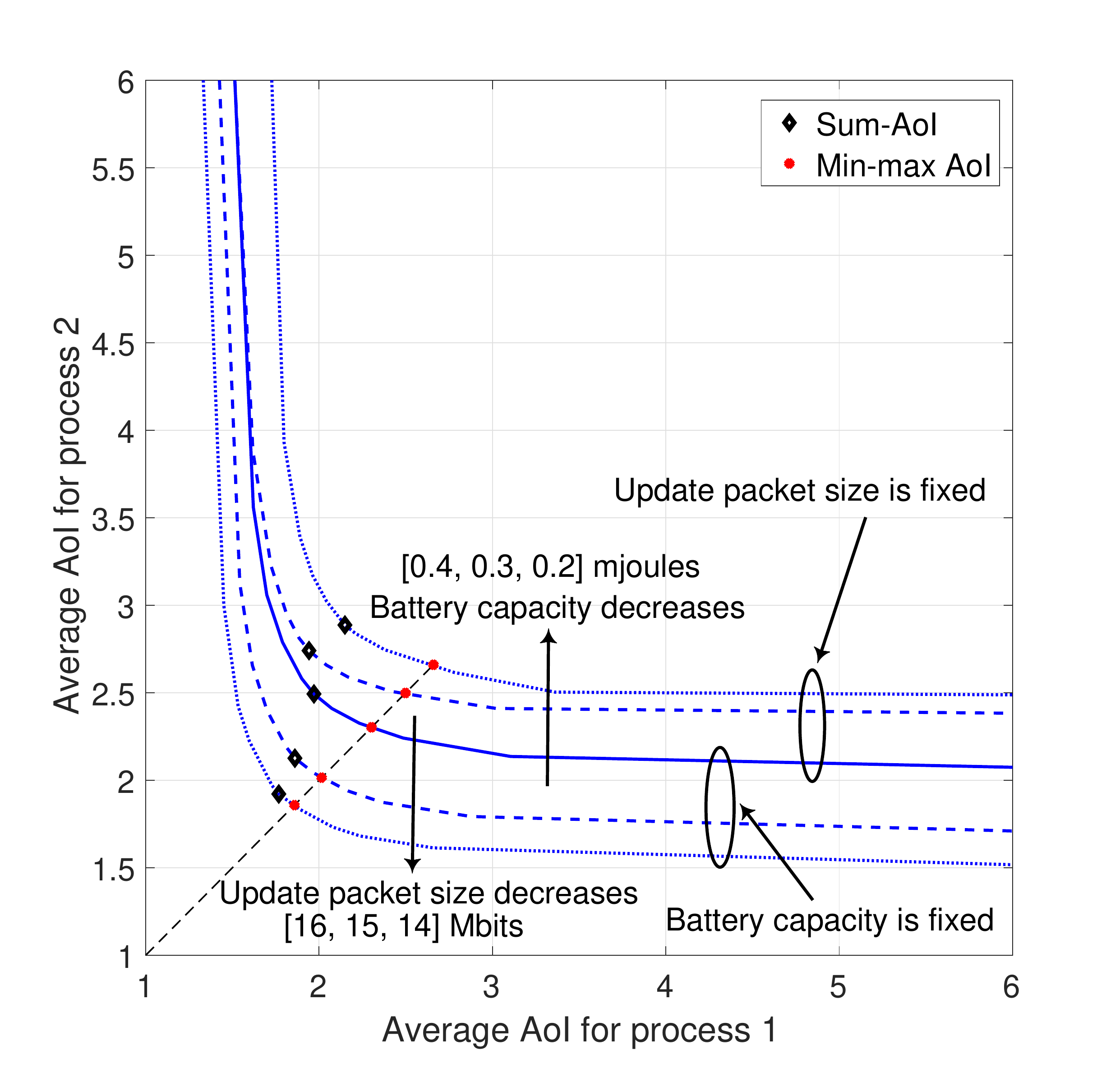}
\caption{Achievable AoI region for the case of two IoT devices. We use $d_1 = 25$ meters, $d_2 = 40$ meters, $A_{1,{\rm max}} = A_{2,{\rm max}} = 6$ and $G = 10$ dB.}
\label{region_battery_packetsize_effect}
\end{figure}

We now consider the case of two IoT devices in Fig. \ref{region_battery_packetsize_effect}, which are observing two arbitrary physical processes and are deployed at different distances from the destination node. Note that the distance between each IoT device and the destination node greatly affects the quality of their channel. This, in turn, has a direct impact on the amounts of energy harvested at each IoT device and the energy required for an update packet transmission about each process, which mainly determine the achievable average AoI values for the two observed processes.
Particularly, Fig. \ref{region_battery_packetsize_effect} shows the impact of size of update packets and battery capacities on the achievable AoI region. It is observed that the achievable AoI region shrinks as the size of the update packets increases or the battery capacities decreases. This is due to the fact that in both cases the allowable number of update packet transmissions by both devices would decrease. Furthermore, as expected, achieving fairness 
comes at the expense of a performance degradation in terms of the achievable sum-AoI value, and the situation worsens as the size of update packets increases or the capacity of batteries decreases. This can be seen by comparing, for each curve, the minimum sum-AoI value achieved by the sum-AoI operating point with the sum-AoI value associated with the min-max operating point. This demonstrates a fundamental trade-off between achieving fairness among both processes and achieving the minimum sum-AoI.
\section{Open Problems and Takeaway Messages}\label{sec:open}
After introducing the concept of AoI and exploring its application to IoT, we will now discuss several key open problems in this area. 
Our hope is that this section will be useful for the new researchers trying to enter this exciting new area. 
\begin{itemize}
\item {\em Characterization of the distribution of AoI.} With the {\em average} AoI and its variants being fairly well studied by now, the next meaningful step is to characterize the distribution of AoI (which is a random process). 
In this direction, a formula for the stationary distribution of AoI has been recently derived in \cite{inoue2018general} in terms of the stationary distributions of the system delay and the Peak AoI. While this is a useful step forward, the analysis is for the case when the AoI increases linearly between any two consecutive time instants of update packet receptions. Therefore, a promising avenue of future work is to extend the analysis of \cite{inoue2018general} to the case where a more general cost of information staleness is considered (e.g., CoUD). Thereafter, it will be useful to study second order properties of the AoI random process, such as the auto correlation function.
\item {\em Network-level analysis of AoI.} Although the average AoI has been well-studied under queueing-theoretic models in the literature \cite{kaul2012real,costa2016age,kosta2017age_mono}, the proposed approaches do not lend themselves to the analysis of large-scale IoT networks. Particularly, such queueing-theoretic analyses do not account for key system effects such as the potential coupling between the locations of deployed IoT devices and their destination nodes, the level of interference at different destination nodes and the density of IoT deployment. 
This necessitates the need for extending the analysis of AoI to large-scale settings using ideas from random spatial models and stochastic geometry. 
\item {\em Low-complexity online schemes.} The computational complexity of solving the MDP encountered in this article using VIA or PIA mainly depends on the number of discrete levels considered for each state variable, i.e., battery, channel gain or AoI. Owing to the generality of our proposed system setup in which actions are taken while taking into account different system parameters as state variables, increasing the number of discrete levels for state variables greatly reduces the feasibility of characterizing the optimal policy in practice. This calls for the need to construct low-complexity schemes suitable for large-scale problems using tools from {\em Approximate dynamic programming} \cite{approx_dynam}.
\item {\em Non-linear RF energy harvesting models.} Although the conventional linear RF energy harvesting model used in this article is highly tractable, it may not always be accurate \cite{boshkovska2015practical}. This is because of the non-linear nature of the RF-to-DC power conversion in practical RF energy harvesting systems, which can lead to significant losses in the amount of harvested energy, thus degrading the performance of RF-powered communication systems. Mathematical treatment of these models in the context of AoI is a useful avenue for future work.
\item {\em Machine learning-based algorithms.} In practice, the destination node may not have complete CSI. In such scenarios, machine learning techniques could be leveraged to learn the state of channel power gains from past experience while dynamically taking decisions. Particularly, {\em reinforcement learning algorithms} could be used to predict the values of unknown parameters and statistically improve the network performance. Investigation of such machine-learning driven techniques in the context of AoI is a fruitful direction of future work.
\end{itemize}
\section{Conclusion}\label{sec:con}
This article provided an accessible introduction to the AoI and its variants. In addition, it also investigated the role of AoI in designing and operating RF-powered freshness-aware IoT networks. We considered a system setup in which IoT devices observe different physical processes and need to transmit status updates about these processes to a destination node. IoT devices were assumed to be solely powered by RF energy transfer by the destination node. We studied the problem of long-term weighted sum-AoI minimization in which we jointly optimized wireless energy transfer by the destination node and scheduling of update packet transmissions from IoT devices. Our results concretely demonstrated that the AoI-optimal and throughput-optimal policies have completely different structures. They also demonstrated a fundamental trade-off between achieving fairness among different processes and achieving the minimum sum-AoI.
%

\bibliographystyle{IEEEtran}
\bibliography{Revisedmanu_v0.7}
\begin{IEEEbiographynophoto}
{Mohamed A. Abd-Elmagid} is a Ph.D. student in the Department of Electrical and Computer Engineering at Virginia Tech. He received his B.Sc. and M.S. degrees in Electrical Engineering from Cairo University, Egypt in 2014 and Nile University, Egypt in 2017, respectively. His research interests include age-of-information, energy harvesting, and machine learning.
\end{IEEEbiographynophoto}
\begin{IEEEbiographynophoto}
{Nikolaos Pappas} (S'07-M'13) received the B.Sc., M.Sc, and Ph.D. degrees in computer science, from the University of Crete, Greece, in 2005, 2007, and 2012, respectively. He received the B.Sc. degree in mathematics from the University of Crete on 2012. He is Associate Professor in mobile telecommunications with the Department of Science and Technology, Link\"{o}ping University, Norrk\"{o}ping, Sweden. His research interests include energy harvesting networks, network-level cooperation, age-of-information, and stochastic geometry. He serves as an Editor for two IEEE journals. 
\end{IEEEbiographynophoto}
\begin{IEEEbiographynophoto} 
{Harpreet S. Dhillon} (S'11-M'13-SM'19) is an Associate Professor of Electrical and Computer Engineering and the Elizabeth and James E. Turner Jr. '56 Faculty Fellow at Virginia Tech. He received the B.Tech. degree from IIT Guwahati in 2008, the M.S. degree from Virginia Tech in 2010, and the Ph.D. degree from the University of Texas at Austin in 2013, all in Electrical Engineering. His research interests include communication theory, wireless networks, stochastic geometry, and machine learning. He is a Clarivate Analytics Highly Cited Researcher and a recipient of five best paper awards. He serves as an Editor for three IEEE journals.
 
\end{IEEEbiographynophoto}
\end{document}